\documentclass[12pt,a4paper]{article}
\usepackage[latin1]{inputenc}
\usepackage{amsmath}
\usepackage{amsfonts}
\usepackage{amssymb}
\usepackage[left=2cm,right=2cm,top=2cm,bottom=2cm]{geometry}
\usepackage{tabularx}
\usepackage{subfig}
\usepackage{wrapfig}
\usepackage{tabularx}
\usepackage{sidecap}
\usepackage{setspace}
\usepackage{graphicx}
\usepackage{multirow}
\usepackage{multicol}
\usepackage{cite}
\RequirePackage[colorlinks,citecolor=blue,urlcolor=blue,linkcolor=blue]{hyperref}
\begin{document}
\title{Analytical models of the Proton Structure Function and the gluon distribution at small $x$ beyond leading order.}
\author{ Luxmi Machahari$^{1,*}$ and D. K. Choudhury$^{2,a}$ %and P. K. Sahariah$^{3,c}$ 
\\ \\ $^1$Department of Physics, Rangapara College, Sonitpur 784505, India. %\\$^2$Physics Academy of North-East, Guwahati-781 014, India. 
\\ $^2$Physics Academy of North-East, Guwahati-781 014, India..  \\% $^3$Department of Physics, Cotton State University, Guwahati-781 001\\ 
$^a$email:dkc.phys@gmail.com \\$^*$email:luxmimachahari@gmail.com  %\\ $^c$email:pksahariah@yahoo.com
}
\date{}
\maketitle
\doublespacing
\begin{abstract}
We incorporate the next-to-leading order (NLO) and the next-to-next-to-leading order (NNLO) effects in the models of the Singlet Structure function $F_2^S(x,t)$ and the gluon distribution $G(x,t)$ using DGLAP equations approximated at small $x$. %To obtain 
Analytical solutions %at each order using the Lagrange's Auxiliary method we need to add additional parameter to soften the $t$-dependence of the strong coupling constant 
both at the next-to leading order (NLO) as well as the next-to-next-to leading order (NNLO) are obtained. %Plausible physical significance of such parameters are discussed. 
We then make comparisons with  exact results.
\end{abstract}

\section{Introduction}
\label{intro}
\label{CA}
Study of Proton structure function at small $x$ is an important area of research in recent years. Recently we have reported an analysis of the proton structure function as well as the gluon distribution at small $x$ using the Taylor approximated \cite{Luxmi,Lux2} DGLAP equations %reported \cite{Luxmi,Lux2} we DGLAP 
\cite{gl,l,d,ap}. % evolution equations for both singlet and gluon structure functions at leading order (LO). The effects of next-to-leading order (NLO) \cite{NLO1} and next-to-next-to-leading order (NNLO) \cite{van1,van2,Moch1,Moch2,Retey} corrected terms in the evolution of parton structure functions are important especially for singlets and gluons at low $x$. The NLO and NNLO QCD approximation for the parton distributions functions of deep inelastic scattering should be included in order to understand perturbative QCD (pQCD). 
The precision of the recent experimental data \cite{h1} demands the correction terms of the splitting function atleast upto NLO \cite{NLO1,NLO2} and preferrably NNLO \cite{Moch1,Moch2,Retey} in DGLAP evolution equations.   %shoud be used in accurate comparison between pQCD theory and experiment and in order to arrive at quantitatively reliable predictions for hard process at present and future energy colliders. Recently the one loop splitting functions \cite{ap,LO1,LO2} and two loop splitting functions \cite{NLO1} governing the DGLAP evolution equations are introduced with a good phenomenological success. To obtain the NNLO approximated parton structure functions one needs the corresponding three-loop splitting functions \cite{G.R}.
 In the present paper we obtain the corresponding $t$ evolution of the structure function both at NLO and NNLO. In order to obtain their analytical forms we use plausible relationship between the singlet and gluon distributions \cite{ly,Akbari,ne} and use Lagrange's Auxiliary method \cite{s} to solve the corresponding first order partial differential equation in $x$ and $t$. In section \ref{CB} we discuss the formalism, sect.\ref{E} is devoted to numerical analysis and lastly in sect.\ref{F} we give our conclusions.% the decoupled $t$ evolutions of the LO DGLAP evolution  equations as reported in \cite{Lux2} are generalised to NLO and NNLO with respect to Taylor series expanding at small $x$ and based on QCD behaviour of the singlet and gluon distribution by Lagrange's Auxiliary method.
\section{Formalism}
\label{CB}
\subsection{Taylor approximated coupled DGLAP equations at small $x$ in NLO and NNLO}
%The proton structure function is $F_2^{ep}(x,Q^2)=\frac{5}{18}F_2^S(x,Q^2)+\frac{3}{18}F_2^{NS}(x,Q^2)$. At very small $x$, we can ignore the non-singlet contribution $F_2^{NS}(x,Q^2)$ to the proton structure function. This leads to only the contribution of the singlet structure function  $F_2^{S}(x,Q^2)$ at LO to it. Thus we can write, $$F_2^{ep}(x,Q^2)=x \sum_{i=1}^{N_f}e_i^2(q_i(x,Q^2)+\bar{q}_i(x,Q^2))$$ where $N_f$ is the flavor number and $e_i$ is the electric charge associated with the $i^{th}$ flavor.
 %At low $x$ and high $Q^2$ the singlet quark distribution is
%essentially driven by the generic instability of
%dominated by the gluon distribution $G(x, Q 2 ) = xg(x, Q^2 )$, where $g(x, Q^2 )$ is the gluon density. It is well explained by the DGLAP equations, which describe perturbative evolution of $G(x, Q^2 )$ and $F_2^S(x, Q^2 )$.
The standard forms of coupled DGLAP evolution equations for the singlet and the gluon distributions are given as \cite{gl,d,ap}
\begin{equation}
\label{E5.1}
\frac{\partial F_2^{S}(x,Q^2)}{\partial \ln Q^2}=\frac{\alpha_s}{2\pi}\int_x^1dz[P_{qq}(z,\alpha_s(Q^2))F_2^{S}(\frac{x}{z},Q^2)+2N_fP_{qg}(z,\alpha_s(Q^2))G(\frac{x}{z},Q^2)]
\end{equation}
\begin{equation}
\label{E5.2}
\frac{\partial G(x,Q^2)}{\partial \ln Q^2}=\frac{\alpha_s}{2\pi}\int_x^1dz[P_{gg}(z,\alpha_s(Q^2))G(\frac{x}{z},Q^2)+P_{gq}(z,\alpha_s(Q^2))F_2^{S}(\frac{x}{z},Q^2)]
\end{equation}
where $N_f$ is the flavor number. $F_2^S(x,Q^2)=x\Sigma(x, Q^2)$ is the singlet structure function, $\Sigma(x, Q^2)=\sum_{i=1}^{N_f}(q_i(x,Q^2)+\bar{q}_i(x,Q^2))$ is the singlet quark distribution and $q_i$ and $\bar{q}_i$ are the quarks and antiquarks of flavour $i$. $G(x, Q^2 ) = xg(x, Q^2 )$ is the gluon distribution function. 
%Where, $F_2^S(x,Q^2)$ and $G(x,Q^2)$ are the singlet and gluon distribution functions respectively and 
The splitting functions $P_{ij}'s$ are Altarelli-Parisi splitting kernels and  are calculable in the perturbative approach to QCD. They have been known, for a long time, at three-loop accuracy Ref.\cite{Moch1,Moch2}, which is the next-to-next-to-leading order (NNLO or N2LO) in the expansion in powers of the strong coupling $\alpha_s$

%The splitting functions $P_{ij}'s$ are the one loop (LO) \cite{ap,LO1,LO2}, two loops (NLO) \cite{NLO2,NLO1} and three loops (NNLO) \cite{Moch1,Moch2,Retey} Altarelli-Parisi splitting kernels given as in the expansion in powers of the strong coupling $\alpha_s$
\begin{equation}
\label{E5.3}
P_{ij}(x,\alpha_s(Q^2))=P_{ij}^{LO}(x)+\frac{\alpha_s(Q^2)}{2\pi}P_{ij}^{NLO}(x)+\left( \frac{\alpha_s(Q^2)}{2\pi}\right)^2 P_{ij}^{NNLO}(x)
\end{equation}
The explicit expressions for $P_{ij}^{LO}$ are as in Ref.\cite{ap,LO1,LO2}, $P_{ij}^{NLO}$ as in Ref. \cite{NLO2,NLO1} and $P_{ij}^{NNLO}$ as in Ref.\cite{Moch1,Moch2,Retey,G.R}. 
%are given in the eqns.(\ref{5A1})-%,(\ref{5A1'}),(\ref{5A1''}),
%(\ref{5A1'''}) and eqns.(\ref{5A3})-%,(\ref{5A3'}),(\ref{5A3''}),
%(\ref{5A3'''}) and (\ref{5A5})-%,(\ref{5A5'}),(\ref{5A5''}),
%(\ref{5A5'''}) respectively of Appendix A for completeness.
%The quark-gluon $(P_{qg})$ and gluon-quark $(P_{qg})$ splitting terms as in eqn.(\ref{E4.1}) and (\ref{E4.2}) are given by $P_{qg}=N_fP_{q_ig}$ and $P_{gq}=P_{gq_i}$ where $P_{q_ig}$ and $P_{gq_i}$ are the flavor-independent splitting functions (Boroun).\\The explicit expressions for $P_{ij}^{LO}$\cite{LO1,LO2} and $P_{ij}^{NLO}$ \cite{NLO1} are given in the Appendix 4.1.
\\
The scale $(Q^2)$ dependence of the strong coupling is controlled by the $\beta-$ function which
can be expressed in perturbative series. The one loop  (LO), two loop (NLO) and three loop (NNLO) solutions of the running coupling constant  $\frac{\alpha_s}{2\pi}$ are respectively \cite{coupling}
\begin{equation}
\label{E5.4}
\frac{\alpha_s^{LO}(t)}{2\pi}=\frac{2}{\beta_0 t}
\end{equation}
\begin{equation}
\label{E5.5}
\frac{\alpha_s^{NLO}(t)}{2\pi}=\frac{2}{\beta_0 t}[1-\frac{\beta_1 \ln t}{\beta_0^2 t}]
\end{equation}
\begin{equation}
\label{E5.6}
\frac{\alpha_s^{NNLO}(t)}{2\pi}=\frac{2}{\beta_0 t}\left[ 1-\frac{\beta_1 \ln t}{\beta_0^2 t}+\frac{1}{(\beta_0 t)^2}\left\lbrace \frac{\beta_1^2}{\beta_0^2}(\ln^2 t-\ln t+1)+\frac{\beta_2}{\beta_0}\right\rbrace \right] 
\end{equation}
Where $\beta_0=\frac{1}{3}(33-2N_f)$, $\beta_1=102-\frac{38}{3}N_f$ and $\beta_2=\frac{2857}{6}-\frac{6673}{18}N_f+\frac{325}{54}N_f^2$ are the one-loop, two-loop and three-loop  corrections respectively to the QCD $\beta$-function and $t$ is defined as $t=\ln(\frac{Q^2}{\Lambda^2})$ and $\Lambda$ is the QCD cut-off scale parameter.\\The Taylor approximation of $F_2^S(\frac{x}{z},Q^2)$ and gluon $G(\frac{x}{z},Q^2)$ upto $O(x)$ \cite{Luxmi,Lux2}
\begin{eqnarray}
\label{E5.7'}
F_2^S(\frac{x}{z},Q^2)=F_2^S(x,Q^2)+x\frac{(1-z)}{z}F_2^S(x,Q^2)\\G(\frac{x}{z},Q^2)=G(x,Q^2)+x\frac{(1-z)}{z}G(x,Q^2)
\end{eqnarray}

Substituting the splitting functions upto NLO and up to NNLO in the DGLAP equations, we obtain the Taylor approximated DGLAP equations \cite{Luxmi,Lux2} for the singlet and gluon distribution as the following\\
\textbf{For NLO:}
\begin{eqnarray}
\label{E5.8'}
\frac{\partial F_2^S(x,t)}{\partial t}=\frac{\alpha_s(t)}{2\pi}\int_x^1 dz[P_{qq}^{LO}F_2^S(x,t)+P_{qq}^{LO}x\frac{(1-z)}{z}\frac{\partial F_2^S(x,t)}{\partial x}+2N_f(P_{qg}^{LO}G(x,t)\nonumber\\+P_{qg}^{LO}x\frac{(1-z)}{z}\frac{\partial G(x,t)}{\partial x}) ] +\left( \frac{\alpha_s(t)}{2\pi}\right) ^2\int_x^1 dz[P_{qq}^{NLO}F_2^S(x,t)\nonumber\\+P_{qq}^{NLO}x\frac{(1-z)}{z}\frac{\partial F_2^S(x,t)}{\partial x}+2N_f(P_{qg}^{NLO}G(x,t)+P_{qg}^{NLO}x\frac{(1-z)}{z}\frac{\partial G(x,t)}{\partial x}) ]
\end{eqnarray}
\begin{eqnarray}
\label{E5.9}
\frac{\partial G(x,t)}{\partial t}=\frac{\alpha_s(t)}{2\pi}\int_x^1 dz[P_{gg}^{LO}G(x,t)+P_{gg}^{LO}x\frac{(1-z)}{z}\frac{\partial G(x,t)}{\partial x}+(P_{gq}^{LO}F_2^S(x,t)\nonumber\\+P_{gq}^{LO}x\frac{(1-z)}{z}\frac{\partial F_2^S(x,t)}{\partial x}) ] +\left( \frac{\alpha_s(t)}{2\pi}\right) ^2\int_x^1 dz[P_{gg}^{NLO}G(x,t)\nonumber\\+P_{gg}^{NLO}x\frac{(1-z)}{z}\frac{\partial G(x,t)}{\partial x}+P_{gq}^{NLO}F_2^S(x,t)+P_{gq}^{NLO}x\frac{(1-z)}{z}\frac{\partial F_2^S(x,t)}{\partial x}]
 \end{eqnarray}
%These are the exact NLO DGLAP expressions with respect to the running coupling constant upto eqn.(\ref{E4.5}) and the splitting functions eqn.(\ref{E5.1}) upto NLO.\\
\textbf{For NNLO:}
\begin{eqnarray}
\label{E5.10}
\frac{\partial F_2^S(x,t)}{\partial t}=\frac{\alpha_s(t)}{2\pi}\int_x^1 dz[P_{qq}^{LO}F_2^S(x,t)+P_{qq}^{LO}x\frac{(1-z)}{z}\frac{\partial F_2^S(x,t)}{\partial x}+2N_f(P_{qg}^{LO}G(x,t)\nonumber\\+P_{qg}^{LO}x\frac{(1-z)}{z}\frac{\partial G(x,t)}{\partial x}) ] +\left( \frac{\alpha_s(t)}{2\pi}\right) ^2\int_x^1 dz[P_{qq}^{NLO}F_2^S(x,t)\nonumber\\+P_{qq}^{NLO}x\frac{(1-z)}{z}\frac{\partial F_2^S(x,t)}{\partial x}+2N_f(P_{qg}^{NLO}G(x,t)+P_{qg}^{NLO}x\frac{(1-z)}{z}\frac{\partial G(x,t)}{\partial x})\nonumber\\+\left( \frac{\alpha_s(t)}{2\pi}\right) ^3\int_x^1 dz[P_{qq}^{NNLO}F_2^S(x,t)+P_{qq}^{NNLO}x\frac{(1-z)}{z}\frac{\partial F_2^S(x,t)}{\partial x}\nonumber\\+2N_f(P_{qg}^{NNLO}G(x,t)+P_{qg}^{NNLO}x\frac{(1-z)}{z}\frac{\partial G(x,t)}{\partial x}) ]
\end{eqnarray}
\begin{eqnarray}
\label{E5.11}
\frac{\partial G(x,t)}{\partial t}=\frac{\alpha_s(t)}{2\pi}\int_x^1 dz[P_{gg}^{LO}G(x,t)+P_{gg}^{LO}x\frac{(1-z)}{z}\frac{\partial G(x,t)}{\partial x}+(P_{gq}^{LO}F_2^S(x,t)\nonumber\\+P_{gq}^{LO}x\frac{(1-z)}{z}\frac{\partial F_2^S(x,t)}{\partial x}) ] +\left( \frac{\alpha_s(t)}{2\pi}\right) ^2\int_x^1 dz[P_{gg}^{NLO}G(x,t)\nonumber\\+P_{gg}^{NLO}x\frac{(1-z)}{z}\frac{\partial G(x,t)}{\partial x}+P_{gq}^{NLO}F_2^S(x,t)+P_{gq}^{NLO}x\frac{(1-z)}{z}\frac{\partial F_2^S(x,t)}{\partial x}\nonumber\\+\left( \frac{\alpha_s(t)}{2\pi}\right) ^3\int_x^1 dz[P_{gg}^{NNLO}G(x,t)+P_{gg}^{NNLO}x\frac{(1-z)}{z}\frac{\partial G(x,t)}{\partial x}\nonumber\\+P_{gq}^{NNLO}F_2^S(x,t)+P_{gq}^{NNLO}x\frac{(1-z)}{z}\frac{\partial F_2^S(x,t)}{\partial x}]
 \end{eqnarray}
% These are the exact NNLO DGLAP  expressions with respect to the running coupling constant upto eqn.(\ref{E5.2}) and the splitting functions eqn.(\ref{E5.1}) upto NNLO.
%\subsection{$t$ evolutions for the singlet and gluon distribution at next-to leading order (NLO) and next-to-next-to leading order (NNLO)}
%\label{C}
Taylor approximated DGLAP eqns.(\ref{E5.8'}), (\ref{E5.9}), (\ref{E5.10}) and (\ref{E5.11}) can be solved only if one assumes plausible analytical relationship between the singlet and gluon distribution. Generally, exact analytical solution of coupled DGLAP equations or an explicit relation between quark and gluon distributions are not possible. Numerical methods are necessary. However, if one assumes that at small $x$ in a certain $Q^2$ range, such analytical form of gluon and quark distribution is possible, the most general form can be written as
\begin{equation}
\label{D1}
G(x,t)=x g(x,t)=x\sum_i^{2N_f}K_iC_i(x,t)\{q_i(x,t)+\bar{q_i}(x,t)\}
\end{equation}
which takes into account the flavor independency of gluon. If the coefficients $K_i$ and $C_i(x,t)$ are flavor independent
then the expression can be written as
\begin{equation}
\label{D1.1}
G(x,t)=KC(x,t)F_2^S(x,t)
\end{equation}
Here, $KC(x, t)$ represents the ratio of the quark and gluon distribution i.e $\frac{G(x,t)}{F_2^S(x,t)}$ and is in general not factorizable in x and t. A purely $t-$independence of the ratio was found to be not true in general \cite{Borou}.
 In conformity with the QCD analysis of Lopez and Yndurain \cite{ly}  and pursued by us later in \cite{Akbari,ne} we use the plausible $t$ dependent relationship %\cite{ly,ne,Luxmi,Lux2} 
between the singlet and the gluon distribution.% compatible with perturbative QCD expectation is 
 \begin{equation}
 \label{E5.12}
 G(x,t)=K t^{\sigma} F_2^S(x,t)
 \end{equation}
 where $K$ and $\sigma$ are fitted from experiments \cite{Lux2}.\\

\textbf{Next-to leading order (NLO):}\\
The above eqns.(\ref{E5.8'}) and (\ref{E5.9}) becomes respectively as
 \begin{eqnarray}
\label{E5.13}
\frac{\partial F_2^S(x,t)}{\partial t}=\frac{\alpha_s(t)}{2\pi}[A_1^S(x,t) F_2^S(x,t)+B_1^S(x,t)\frac{\partial F_2^S(x,t)}{\partial x} ]+\left( \frac{\alpha_s(t)}{2\pi}\right) ^2[C_1^S(x,t) F_2^S(x,t)\nonumber\\+ D_1^S(x,t)\frac{\partial F_2^S(x,t)}{\partial x}]
\end{eqnarray}
\begin{eqnarray}
\label{E5.14}
\frac{\partial F_2^S(x,t)}{\partial t}=\frac{\alpha_s(t)}{2\pi}[A_2^S(x,t) F_2^S(x,t)+B_2^S(x,t)\frac{\partial F_2^S(x,t)}{\partial x} ]+\left( \frac{\alpha_s(t)}{2\pi}\right) ^2[C_2^S(x,t) F_2^S(x,t)\nonumber\\+ D_2^S(x,t)\frac{\partial F_2^S(x,t)}{\partial x}]
 \end{eqnarray}
 where $A_{1,2}^S(x,t), B_{1,2}^S(x,t),C_{1,2}^S(x,t),D_{1,2}^S(x,t)$ are the integrals over splitting functions as given in the eqns. (\ref{B1.1})-(\ref{B2.4}) of Appendix A. The Lagrange's auxiliary method \cite{s} can be applied to solve eqns.(\ref{E5.13}) and (\ref{E5.14}) analytically only if the $t$ evolution of the strong coupling constant at NLO $(\alpha_s^{NLO}(t))^2$ can be linearised. 
 %To proceed further with Lagrange's method we linearize $T^2(t)$ as $T_1T(t)$, where 
 Defining $T(t)=\frac{\alpha_s(t)}{2\pi}$ ,  we linearise it to be $$\left( \frac{\alpha_s(t)}{2\pi}\right) ^2=T_1T(t)$$$$\implies T^2(t)=T_1T(t)$$ \cite{Atri2,PKD3,PKD4, Saiful1,JK2}, where $T_1$ is a parameter to be determined from the particular  range of $Q^2$ range under study.
 Following Lagrange$'$s method \cite{s} to solve eqns. (\ref{E5.13}) and (\ref{E5.14}), we put them in the form as,
\begin{eqnarray}
\label{E5.15}
Q_1^{NLO}(x,t)\frac{\partial F_2^S(x,t)}{\partial t}+P_1^{NLO}(x,t)\frac{\partial F_2^S(x,t)}{\partial x}=R_1^{NLO}(x,t)F_2^S(x,t)
 \end{eqnarray} 
 \begin{eqnarray}
\label{E5.16}
Q_2^{NLO}(x,t)\frac{\partial F_2^S(x,t)}{\partial t}+P_2^{NLO}(x,t)\frac{\partial F_2^S(x,t)}{\partial x}=R_2^{NLO}(x,t)F_2^S(x,t)
\end{eqnarray} 
 Where,
 \begin{eqnarray}
\label{E5.17}
Q_1^{NLO}(x,t)=1\\P_1^{NLO}(x,t)=-T(t)[B_1^S(x,t)+T_1 D_1^S(x,t)]\\R_1^{NLO}(x,t)=T(t)[A_1^S(x,t)+T_1 C_1^S(x,t)]
 \end{eqnarray}
 and,
 \begin{eqnarray}
\label{E5.18}
Q_2^{NLO}(x,t)=1\\P_2^{NLO}(x,t)=-T(t)[B_2^S(x,t)+T_1 D_2^S(x,t)]\\R_2^{NLO}(x,t)=T(t)[A_2^S(x,t)+T_1 C_2^S(x,t)]
 \end{eqnarray}

% Taking $\hat{T}(t)=\frac{\alpha_s(t)}{2\pi}$ and $\hat{T}^2(t)$ is linearised through the ansatz $\hat{T}^2(t)=\hat{T}_0\hat{T}(t)$ 
 The Lagrange's equations (\ref{E5.15}) and (\ref{E5.16}) is obtained from the solutions of the auxiliary equation
 \begin{equation}
 \label{E5.19}
 \frac{dx}{P_{1,2}^{NLO}(x,t)}=\frac{dt}{1}=\frac{dF_2^S(x,t)}{R_{1,2}^{NLO}(x,t)F_2^S(x,t)}
 \end{equation}
 The general solution of equations (\ref{E5.15}) and (\ref{E5.16}) is given by
  \begin{equation}
 \label{E5.20}
 f(u_{1,2}^{NLO},v_{1,2}^{NLO})=0
 \end{equation}
 Where $f(u_{1,2}^{NLO},v_{1,2}^{NLO})$ is arbitrary function of $u_{1,2}^{NLO},v_{1,2}^{NLO}$ are defined below in eqns.(\ref{E5.21}) and (\ref{E5.22}). Let  $u_{1,2}^{NLO}(x,t)=C_1^{'}$ and $v_{1,2}^{NLO}(x,t,F_2^S(x,t))=D_1^{'}$ be two independent solutions of eqn.(\ref{E5.19}). Solving eqn.(\ref{E5.20}) we obtain
  \begin{eqnarray}
 \label{E5.21}
 u_{1,2}^{NLO}(x,t)=t X_{1,2}^{NLO}(x,t)\\v_{1,2}^{NLO}(x,t,F_2^S(x,t))=F_2^S(x,t)Y_{1,2}^{NLO}(x,t)
 \end{eqnarray}
 Where,
  \begin{equation}
 \label{E5.22}
 X_{1,2}^{NLO}(x,t)=t^{\frac{b}{t}}e^{\frac{b}{t}}\exp\left[ \frac{1}{a}\int\frac{dx}{B_{1,2}^S(x,t)+T_1D_{1,2}^S(x,t)}\right] 
 \end{equation}
 \begin{equation}
 \label{E5.23}
 Y_{1,2}^{NLO}(x,t)=\exp\left[ \int\frac{A_{1,2}^S(x,t)+T_1 C_{1,2}^S(x,t)}{B_{1,2}^S(x,t)+T_1D_{1,2}^S(x,t)}dx\right] 
 \end{equation}
 and $a=\frac{2}{\beta_0}$ ; $b=\frac{\beta_1}{\beta_0^2}$.% The functions $A_{1,2}^S,B_{1,2}^S,C_{1,2}^S,D_{1,2}^S$ are as given in the appendix A.
\\The linear combination of $u_{1,2}^{NLO}(x,t)$ and $v_{1,2}^{NLO}(x,t,F_2^S(x,t))$ in $F_2^S(x,t)$ as in \cite{Luxmi} 
\begin{equation}
u_{1,2}^{NLO}+\alpha v_{1,2}^{NLO}=\beta^{NLO}
\end{equation} gives
 \begin{equation}
 \label{E5.24}
 F_2^{S(I,II),NLO}(x,t)=\frac{1}{\alpha Y_{1,2}^{NLO}(x,t)}[\beta-tX_{1,2}^{NLO}(x,t) ]
 \end{equation}
 $F_2^{S(I),NLO}(x,t)$ and $F_2^{S(II),NLO}(x,t)$  are the solutions of eqns.(\ref{E5.15}) and (\ref{E5.16}) respectively.
 Using the boundary condition at certain $t=t_0,$
\begin{equation}
 \label{E5.25}
 F_2^{S(I),NLO}(x,t_0)=F_2^{S(II),NLO}(x,t_0)=F_2^{S}(x,t_0)
 \end{equation}
 We obtain two alternative $t-$ evolution equations for Singlet distribution analytically at NLO as
 \begin{equation}
 \label{E5.26}
 F_2^{S(I),NLO}(x,t)=F_2^{S}(x,t_0)\frac{t}{t_0}\left( \frac{Y_{1}^{NLO}(x,t_0)}{ Y_{1}^{NLO}(x,t)}\right) \left[ \frac{X_{1}^{NLO}(x,t)-\frac{\beta}{t}}{X_{1}^{NLO}(x,t_0)-\frac{\beta}{t_0}} \right] 
 \end{equation}
 \begin{equation}
 \label{E5.27}
 F_2^{S(II),NLO}(x,t)=F_2^{S}(x,t_0)\frac{t}{t_0}\left( \frac{Y_{2}^{NLO}(x,t_0)}{ Y_{2}^{NLO}(x,t)}\right) \left[ \frac{X_{2}^{NLO}(x,t)-\frac{\beta}{t}}{X_{2}^{NLO}(x,t_0)-\frac{\beta}{t_0}} \right] 
 \end{equation}
 With ratio
 \begin{equation}
 \label{E5.28}
 R^{NLO}(x,t)=\frac{F_2^{S(I),NLO}(x,t)}{F_2^{S(II),NLO}(x,t)} =\frac{\left( \frac{Y_{1}^{NLO}(x,t_0)}{ Y_{1}^{NLO}(x,t)}\right) \left[ \frac{X_{1}^{NLO}(x,t)-\frac{\beta}{t}}{X_{1}^{NLO}(x,t_0)-\frac{\beta}{t_0}} \right] }{\left( \frac{Y_{2}^{NLO}(x,t_0)}{ Y_{2}^{NLO}(x,t)}\right) \left[ \frac{X_{2}^{NLO}(x,t)-\frac{\beta}{t}}{X_{2}^{NLO}(x,t_0)-\frac{\beta}{t_0}} \right] }
 \end{equation}
 which is  not equal to unity in general as in LO \cite{Luxmi,Lux2}. Even if  the factor $\beta$ vanishes, because of the $t$ dependence of the functions $X_{1}^{NLO}(x,t), Y_{1}^{NLO}(x,t),X_{2}^{NLO}(x,t), Y_{2}^{NLO}(x,t)$, the ratio will not be identity \cite{Luxmi,Lux2}.\\
 
 The integral functions of $X_{1}^{NLO}(x,t),Y_{1}^{NLO}(x,t),X_{2}^{NLO}(x,t),Y_{2}^{NLO}(x,t)$ occuring in eqns. (\ref{E5.26}) and (\ref{E5.27})   are as follows:
\small
 \begin{equation}
 \label{D1}
 X_{1}^{NLO}(x,t)=t^{\frac{b}{t}}e^{\frac{b}{t}}\exp\left[ \frac{1}{a} \int \frac{dx}{a_1(K,t,\sigma)+b_1(K,t,\sigma)x+c_1(K,t,\sigma)x \ln(1/x)}\right]
 \end{equation}
  \begin{equation}
  \label{D1'}
 Y_{1}^{NLO}(x,t)=\exp\left[ \int \frac{(a_{2}(K,t,\sigma)+b_{2}(K,t,\sigma)x+d_{2}(K,t,\sigma) \ln(1/x))dx}{a_1(K,t,\sigma)+b_1(K,t,\sigma)x+c_1(K,t,\sigma)x \ln(1/x)}\right] 
  \end{equation}
  \begin{equation}
  \label{D1''}
  X_{2}^{NLO}(x,t)=t^{\frac{b}{t}}e^{\frac{b}{t}}\exp\left[\frac{1}{a} \int \frac{dx}{a_3(K,t,\sigma)+b_3(K,t,\sigma)x+c_3(K,t,\sigma)x \ln(1/x)+d_3(K,t,\sigma) \ln(1/x)}\right]
  \end{equation}
 \begin{equation}
 \label{D1'''}
 Y_{2}^{NLO}(x,t)=\exp\left[ \int \frac{(a_{4}(K,t,\sigma)+b_{4}(K,t,\sigma)x+d_{4}(K,t,\sigma) \ln(1/x))dx}{a_3(K,t,\sigma)+b_3(K,t,\sigma)x+c_3(K,t,\sigma)x \ln(1/x)+d_3(K,t,\sigma)\ln(1/x)}\right]
 \end{equation}
 The coefficients $a_1(K,t,\sigma), b_1(K,t,\sigma),c_1(K,t,\sigma), a_2(K,t,\sigma),b_2(K,t,\sigma), d_2(K,t,\sigma),a_3(K,t,\sigma),\\b_3(K,t,\sigma),c_3(K,t,\sigma),d_3(K,t,\sigma),a_4(K,t,\sigma),b_4(K,t,\sigma),d_4(K,t,\sigma)$ are as given in the Appendix B.
\textbf{Next-to-next-to leading order (NNLO):}\\
 For next-to-next-to leading order, substituting eqn.(\ref{E5.12}) in the  the evolution equations for the singlet and gluon distributions eqns.(\ref{E5.10}) and (\ref{E5.11}) respectively becomes
 \begin{eqnarray}
\label{E5.30}
\frac{\partial F_2^S(x,t)}{\partial t}=\frac{\alpha_s(t)}{2\pi}[A_1^S(x,t) F_2^S(x,t)+B_1^S(x,t)\frac{\partial F_2^S(x,t)}{\partial x} ]+\left( \frac{\alpha_s(t)}{2\pi}\right) ^2[C_1^S(x,t) F_2^S(x,t)\nonumber\\+ D_1^S(x,t)\frac{\partial F_2^S(x,t)}{\partial x}+\left( \frac{\alpha_s(t)}{2\pi}\right) ^3[L_1^S(x,t) F_2^S(x,t)+ M_1^S(x,t)\frac{\partial F_2^S(x,t)}{\partial x}]
\end{eqnarray}
\begin{eqnarray}
\label{E5.31}
\frac{\partial F_2^S(x,t)}{\partial t}=\frac{\alpha_s(t)}{2\pi}[A_2^S(x,t) F_2^S(x,t)+B_2^S(x,t)\frac{\partial F_2^S(x,t)}{\partial x} ]+\left( \frac{\alpha_s(t)}{2\pi}\right) ^2[C_2^S(x,t) F_2^S(x,t)\nonumber\\+ D_2^S(x,t)\frac{\partial F_2^S(x,t)}{\partial x}+\left( \frac{\alpha_s(t)}{2\pi}\right) ^3[L_2^S(x,t) F_2^S(x,t)+ M_2^S(x,t)\frac{\partial F_2^S(x,t)}{\partial x}]
 \end{eqnarray}
 %$A_{1,2}^S(x,t),B_{1,2}^S(x,t),C_{1,2}^S(x,t),D_{1,2}^S(x,t)$, are as earlier given in the eqns. (\ref{B1.1})-(\ref{B2.4}) and 
 Where, $L_{1,2}^S(x,t), M_{1,2}^S(x,t)$ are as given in the eqns.(\ref{B1.5})-(\ref{B2.6}) of Appendix A.\\
 
 To proceed further and solve the above two eqns.(\ref{E5.30}) and (\ref{E5.31})by Lagrange's method, we need to linearize the cubic term of the strong coupling constant at NNLO. %Defining  $\hat{T}(t)=\frac{\alpha_s^{NNLO}(t)}{2\pi}$,   %$\hat{T}^2(t)$ and $\hat{T}^3(t)$ are 
 We linearise through the ansatz %$\hat{T}^2(t)=T_2\hat{T}(t)$ and 
 $T^3(t)={T_2}^2 T(t)$ , where $T(t)=\frac{\alpha_s(t)}{2\pi}$\cite{Atri2,PKD3,PKD4,Saiful1,JK2} and  $T_2$ is a suitable parameter to be determined from the particular  range of $Q^2$ range under study.
% Equations (\ref{E5.5}) and (\ref{E5.6}) can be written in the form as,
%\begin{eqnarray}
%\label{E5.7}
%Q_1^{NNLO}\frac{\partial F_2^S(x,t)}{\partial t}+P_1^{NNLO}(x,t)\frac{\partial F_2^S(x,t)}{\partial x}=R_1^{NNLO}(x,t)F_2^S(x,t)
% \end{eqnarray} 
% \begin{eqnarray}
%\label{E5.8}
%Q_2^{NNLO}\frac{\partial F_2^S(x,t)}{\partial t}+P_2^{NNLO}(x,t)\frac{\partial F_2^S(x,t)}{\partial x}=R_2^{NNLO}(x,t)F_2^S(x,t)
%\end{eqnarray} 
% Where,
Solving the following Lagrange's equations
 \begin{eqnarray}
\label{E5.7}
Q_1^{NNLO}\frac{\partial F_2^S(x,t)}{\partial t}+P_1^{NNLO}(x,t)\frac{\partial F_2^S(x,t)}{\partial x}=R_1^{NNLO}(x,t)F_2^S(x,t)
 \end{eqnarray} 
 \begin{eqnarray}
\label{E5.8}
Q_2^{NNLO}\frac{\partial F_2^S(x,t)}{\partial t}+P_2^{NNLO}(x,t)\frac{\partial F_2^S(x,t)}{\partial x}=R_2^{NNLO}(x,t)F_2^S(x,t)
\end{eqnarray} 
 Where,\begin{eqnarray}
\label{E5.35}
Q_{1,2}^{NNLO}(x,t)=1\\P_{1,2}^{NNLO}(x,t)=-T(t)[B_{1,2}^S(x,t)+T_1 D_{1,2}^S(x,t)+T_2^2M_{1,2}^S(x,t)]\\R_{1,2}^{NNLO}(x,t)=T(t)[A_{1,2}^S(x,t)+T_1 C_{1,2}^S(x,t)+T_2^2L_{1,2}^S(x,t)]
 \end{eqnarray}
% and,
% \begin{eqnarray}
%\label{E5.36}
%Q_2^{NNLO}(x,t)=1\\P_2^{NNLO}(x,t)=-\hat{T}(t)[B_2^S(x,t)+T_2 D_2^S(x,t)+T_2^2M_2^S(x,t)]\\R_2^{NNLO}(x,t)=\hat{T}(t)[A_2^S(x,t)+T_2 C_2^S(x,t)+T_2^2L_2^S(x,t)]
% \end{eqnarray}
%
% The Lagrange's equations (\ref{E5.7}) and (\ref{E5.8}) is obtained from the solutions of the auxiliary equation
% \begin{equation}
% \label{E5.11}
% \frac{dx}{P_{1,2}^{NNLO}(x,t)}=\frac{dt}{1}=\frac{dF_2^S(x,t)}{R_{1,2}^{NNLO}(x,t)F_2^S(x,t)}
% \end{equation}
% The general solution of equations (\ref{E5.7}) and (\ref{E5.8}) is given by
%  \begin{equation}
% \label{E5.12}
% f(u_{1,2}^{NNLO},v_{1,2}^{NNLO})=0
% \end{equation}
% Where $f(u_{1,2}^{NNLO},v_{1,2}^{NNLO})$ is arbitrary function of $u_{1,2}^{NNLO},v_{1,2}^{NNLO}$. Let  $u_{1,2}^{NNLO}(x,t)=C_1^{''}$ and $v_{1,2}^{NNLO}(x,t,F_2^S(x,t))=D_1^{''}$ are two independent solutions of eqn.(\ref{E5.11}). Solving eqn.(\ref{E5.11}) we obtain
%Solving by the same method of Lagrange's Auxiliary using eqns.(\ref{E5.30}) and (\ref{E5.31}) 
we obtain,
  \begin{eqnarray}
 \label{E5.32}
 u_{1,2}^{NNLO}(x,t)=t X_{1,2}^{NNLO}(x,t)\\v_{1,2}^{NNLO}(x,t,F_2^S(x,t))=F_2^S(x,t)Y_{1,2}^{NNLO}(x,t)
 \end{eqnarray}
 Where,
  \begin{equation}
 \label{E5.33}
 X_{1,2}^{NNLO}(x,t)=t^{\frac{b}{t}}e^{(\frac{b}{t}-\frac{(b^2 \ln^2 t+b^2+c)}{2t^2})}\exp\left[ \frac{1}{a}\int\frac{dx}{B_{1,2}^S(x,t)+T_1D_{1,2}^S(x,t)+T_2^2M_{1,2}^S(x,t)}\right] 
 \end{equation}
 \begin{equation}
 \label{E5.34}
 Y_{1,2}^{NNLO}(x,t)=\exp\left[\int\frac{A_{1,2}^S(x,t)+T_1 C_{1,2}^S(x,t)+T_2^2L_{1,2}^S(x,t)}{B_{1,2}^S(x,t)+T_1D_{1,2}^S(x,t)+T_2^2M_{1,2}^S(x,t)}dx\right] 
 \end{equation}
Where $a=\frac{2}{\beta_0}$ ; $b=\frac{\beta_1}{\beta_0^2}$ as introduced after eqn.(\ref{E5.23}) and $c=\frac{\beta_2}{\beta_0^3}$.

% Taking $\hat{T}(t)=\frac{\alpha_s(t)^{NNLO}}{2\pi}$ and $\hat{T}^2(t)$ and $\hat{T}^3(t)$ are linearised through the ansatz $\hat{T}^2(t)=T_2\hat{T}(t)$ and $\hat{T}^3(t)=T_3\hat{T}(t)$ respectively. Thus effectively, $T_2=\hat{T}(t)$ and $T_3=\hat{T}^2(t)$. %The functions $A_{1,2}^S,B_{1,2}^S,C_{1,2}^S,D_{1,2}^S$ are as given in the appendix 4.1.
%\\The linear combination of $u_{1,2}^{NNLO}(x,t)$ and $v_{1,2}^{NNLO}(x,t,F_2^S(x,t))$ in $F_2^S(x,t)$ gives
% \begin{equation}
% \label{E4.21}
% F_2^{S(I,II)NNLO}(x,t)=\frac{1}{\alpha Y_{1,2}^{NNLO}(x,t)}[\beta-tX_{1,2}^{NNLO}(x,t) ]
% \end{equation}
% $F_2^{S(I),NNLO}(x,t)$ and $F_2^{S(II),NNLO}(x,t)$  are the solutions of eqns.(\ref{E5.5}) and (\ref{E5.6}) respectively.
% Using the boundary condition at $t=t_0$
%\begin{equation}
% \label{E4.22}
% F_2^{S(I),NNLO}(x,t)=F_2^{S(II),NNLO}(x,t)=F_2^{S}(x,t_0)
% \end{equation}
The linear combination of $u_{1,2}^{NNLO}(x,t)$ and $v_{1,2}^{NNLO}(x,t,F_2^S(x,t))$ in $F_2^S(x,t)$ as in \cite{Luxmi} 
\begin{equation}
u_{1,2}^{NNLO}+\alpha v_{1,2}^{NNLO}=\beta^{NNLO}
\end{equation} gives two alternative $t$ evolution euation for Singlet distribution in NNLO as
 \begin{equation}
 \label{E5.37}
 F_2^{S(I),NNLO}(x,t)=F_2^{S}(x,t_0)\frac{t}{t_0}\left( \frac{Y_{1}^{NNLO}(x,t_0)}{ Y_{1}^{NNLO}(x,t)}\right) \left[ \frac{X_{1}^{NNLO}(x,t)-\frac{\beta}{t}}{X_{1}^{NNLO}(x,t_0)-\frac{\beta}{t_0}} \right] 
 \end{equation}
 \begin{equation}
 \label{E5.38}
 F_2^{S(II),NNLO}(x,t)=F_2^{S}(x,t_0)\frac{t}{t_0}\left( \frac{Y_{2}^{NNLO}(x,t_0)}{ Y_{2}^{NNLO}(x,t)}\right) \left[ \frac{X_{2}^{NNLO}(x,t)-\frac{\beta}{t}}{X_{2}^{NNLO}(x,t_0)-\frac{\beta}{t_0}} \right] 
 \end{equation}

 With ratio
 \begin{equation}
 \label{E5.39}
 R^{NNLO}(x,t)=\frac{F_2^{S(I),NNLO}(x,t)}{F_2^{S(II),NNLO}(x,t)} =\frac{\left( \frac{Y_{1}^{NNLO}(x,t_0)}{ Y_{1}^{NNLO}(x,t)}\right) \left[ \frac{X_{1}^{NNLO}(x,t)-\frac{\beta}{t}}{X_{1}^{NNLO}(x,t_0)-\frac{\beta}{t_0}} \right] }{\left( \frac{Y_{2}^{NNLO}(x,t_0)}{ Y_{2}^{NNLO}(x,t)}\right) \left[ \frac{X_{2}^{NNLO}(x,t)-\frac{\beta}{t}}{X_{2}^{NNLO}(x,t_0)-\frac{\beta}{t_0}} \right] }
 \end{equation}
 Which is  not equal to unity in general as discussed in LO \cite{Luxmi,Lux2} and NLO because of the $t$ dependence of $X_1^{NNLO}(x,t)$ and $X_2^{NNLO}(x,t)$ even if $\beta$ is set to zero. It is to be noted that the term $\beta$ occuring in eqns.(\ref{E5.26}), (\ref{E5.27}) and (\ref{E5.37}), (\ref{E5.38}) at LO, NLO and NNLO may not be in general identical i.e $\beta^{LO}\neq \beta^{NLO}\neq\beta^{NNLO}$.

The equations (\ref{E5.26}), (\ref{E5.27}) and (\ref{E5.37}), (\ref{E5.38}) are our main analytical expressions for singlet distributions in NLO and NNLO respectively. %It is therefore important to see if under special conditions the two evolutions will be identical so that the ratio will be equal to unity both for NLO and NNLO.
 The explicit expressions for $X_{1,2}^{NNLO}(x,t)$, $Y_{1,2}^{NNLO}(x,t)$ occuring eqns.(\ref{E5.37}) and (\ref{E5.38}) are as follows.
\begin{equation}
  \label{D2}
 X_{1}^{NNLO}(x,t)=t^{\frac{b}{t}}e^{\frac{b}{t}-\frac{b^2(\ln t)^2+b^2+c}{2t^2}}\exp\left[ \frac{1}{a} \int \frac{dx}{a_5(K,t,\sigma)+b_5(K,t,\sigma)x+c_5(K,t,\sigma)x \ln(1/x)}\right]  
  \end{equation}
  \begin{equation}
  \label{D2'}
 Y_{1}^{NNLO}(x,t)=\exp\left[ \int \frac{(a_{6}(K,t,\sigma)+b_{6}(K,t,\sigma)x+d_{6}(K,t,\sigma) \ln(1/x))dx}{a_5(K,t,\sigma)+b_5(K,t,\sigma)x+c_5(K,t,\sigma)x \ln(1/x)}\right] 
  \end{equation}
  \begin{equation}
  \label{D2''}
X_{2}^{NNLO}(x,t)=t^{\frac{b}{t}}e^{\frac{b}{t}-\frac{b^2(\ln t)^2+b^2+c}{2t^2}}\exp\left[ \frac{1}{a} \int \frac{dx}{a_7(K,t,\sigma)+b_7(K,t,\sigma)x+c_7(K,t,\sigma)x \ln(1/x)+d_7(K,t,\sigma) \ln(1/x)}\right]  
  \end{equation}
 \begin{equation}
 \label{D2'''}
  Y_{2}^{NNLO}(x,t)=\exp\left[\int \frac{(a_{8}(K,t,\sigma)+b_{8}(K,t,\sigma)x+d_{8}(K,t,\sigma) \ln(1/x))dx}{a_7(K,t,\sigma)+b_7(K,t,\sigma)x+c_7(K,t,\sigma)x \ln(1/x)+d_7(K,t,\sigma)\ln(1/x)}\right]
 \end{equation} 
  The coefficients $a_5(K,t,\sigma), b_5(K,t,\sigma),c_5(K,t,\sigma), a_6(K,t,\sigma),b_6(K,t,\sigma), d_6(K,t,\sigma),a_7(K,t,\sigma),b_7(K,t,\sigma),\\c_7(K,t,\sigma),d_7(K,t,\sigma),a_8(K,t,\sigma),b_8(K,t,\sigma),d_8(K,t,\sigma)$ are as given in the Appendix B.
A structure of these functions indicate that they are not analytically solvable.% In the next subsection we will explore the possibility of obtaining their analytical structure making plausible approximations. % However at  ultra small $x$ limit if we assume only the dominant logarithmic term in the denominator  compared to other terms, one can obtain %irrespective of the values of $K,\sigma$ in the functional coefficients only then we can obtain 
\subsection{Approximate analytical expressions for structure functions %$F_2^{S(I)}(x,t)$ and $F_2^{S(II)}(x,t)$
}
\subsubsection{ At Leading order (LO)}
\label{1}
For numerical analysis of the leading order we use our previous result as given in the following equations of Ref\cite{Lux2} 
\begin{equation}
\label{E5.38a}
F_2^{S(I),LO}(x,t)=F_2^S(x,t_0) \left( \frac{t}{t_0}\right)  \frac{Y_1^{LO}(x,t_0)}{Y_1^{LO}(x,t)}\left[\frac{X_1^{LO}(x,t)-\frac{\beta}{t}}{X_1^{LO}(x,t_0)-\frac{\beta}{t_0}} \right] 
\end{equation} 

\begin{equation}
\label{E5.38b}
F_2^{S(II),LO}(x,t)=F_2^S(x,t_0) \left( \frac{t}{t_0}\right) \frac{Y_2^{LO}(x,t_0)}{Y_2^{LO}(x,t)}\left[\frac{X_2^{LO}(x,t)-\frac{\beta}{t}}{X_2^{LO}(x,t_0)-\frac{\beta}{t_0}} \right] 
\end{equation} 
Where the analytical forms of $X_1^{LO}(x,t), Y_1^{LO}(x,t), X_2^{LO}(x,t)$ and $Y_2^{LO}(x,t)$ are as the following.
\begin{equation}
\label{E5.38c}
X_1^{LO}(x,K,t,\sigma)=\exp\left[ -\frac{\ln(\ln\frac{1}{x})}{A_f (2+ \frac{3N_f}{2}Kt^{\sigma})} \right] 
\end{equation}
\begin{equation}
\label{E5.38d}
Y_1^{LO}(x,K,t,\sigma)=\exp[-\frac{1}{(2+\frac{3}{2}N_fKt^{\sigma})}(\int\frac{(2- \frac{3}{2}N_fKt^{\sigma})}{\ln\frac{1}{x}}dx+\int\frac{N_fKt^{\sigma}}{x\ln\frac{1}{x}}dx)]\nonumber\\
\end{equation}
\begin{equation}
\label{E5.38e}
X_2^{LO}(x,K,t,\sigma)=\exp\left[ \frac{Kt^{\sigma}}{9A_f}\frac{x}{(Kt^{\sigma}+\frac{4}{9})}\right] 
\end{equation}
\begin{equation}
\label{E5.38f}
Y_2^{LO}(x,K,t,\sigma)=\exp\left[\frac{-Kt^{\sigma}(9+\frac{4}{Kt^{\sigma}})}{9(Kt^{\sigma}+\frac{4}{9})}x \ln\frac{1}{x} \right] 
\end{equation}
And $A_f=\frac{4}{3\beta_0}$\\
From the phenomenological observation of both $F_2^{S(I),LO}(x,t)$ and $F_2^{S(II),LO}(x,t)$ as in Ref\cite{Lux2}, we obtain $F_2^{S(I),LO}(x,t)$ to be more preferred than the other because of the range of validity in $x$ and $Q^2$ \cite{Luxmi} and hence we make the further numerical analysis of NLO and NNLO only with the first evolutions $F_2^{S(I),NLO}(x,t)$ and $F_2^{S(I),NNLO}(x,t)$. %The best value of $\beta, K$ and $\sigma$  obtained were $-1.016$, $1.856$ and $0.227$ respectively. \\
%For comparison with NLO and NNLO we also find the corresponding $Q^2$ value, where NLO and NNLO approximations are valid.
\subsubsection{ At next-to-leading order (NLO) and at next-to-next-to-leading order (NNLO)}
\label{D}
The analytical forms of the structure functions as in eqns. (\ref{E5.26}) at NLO and (\ref{E5.37}) at NNLO are possible only if we make additional assumptions below whose validity will be tested subsequently.

We observe that at NLO, among the three functions $a_1(K,t,\sigma)$, $b_1(K,t,\sigma)$ and $c_1(K,t,\sigma)$ arising in eqn.(\ref{E5.26}) as given in Appendix B, for a given $K,t, \sigma$, the function $a_1(K,t,\sigma)$ is smaller compared to $b_1(K,t,\sigma)$ and $c_1(K,t,\sigma)$, in the denominators of $X_1^{NLO}(x,t)$ and $Y_1^{NLO}(x,t)$ of eqns.(\ref{D1}) and (\ref{D1'}) respectively. Under this assuumption $X_1^{NLO}(x,t)$ and $Y_1^{NLO}(x,t)$ are obtained as the following.
 \begin{equation}
 \label{E5.43n}
 X_{1}^{NLO}(x,t)=t^{\frac{b}{t}}e^{\frac{b}{t}}\exp\left[ \frac{1}{a}\left( \frac{-1}{b_1}\log{\frac{1}{x}}+\frac{c_1}{b_1^{2}}\frac{(\log{\frac{1}{x})^2}}{2}\right) \right] 
 \end{equation}
 \begin{equation}
 \label{E5.43o}
 Y_{1}^{NLO}(x,t)=\exp\left[ \left( \frac{b_2}{b_1}-\frac{b_2c_1}{b_1^{2}}\right)x-\frac{a_2}{b_1} \log\frac{1}{x}-\left( \frac{d_2}{b_1}-\frac{a_2c_1}{b_1^{2}}\right)\frac{(\log\frac{1}{x})^2}{2}+\frac{d_2c_1}{b_1^{2}}\frac{(\log\frac{1}{x})^3}{3}-\frac{b_2c_1}{b_1^{2}}x\log\frac{1}{x}\right] 
 \end{equation}\\
 Where $a=\frac{2}{\beta_0}$ ; $b=\frac{\beta_1}{\beta_0^2}$ as introduced after eqn.(\ref{E5.23}) and $c=\frac{\beta_2}{\beta_0^3}$. and the coeffecients $b_1, c_1,a_2,b_2,d_2$ are as given in the Appendix B.
 
 Hence
 \begin{equation}
 \label{E5.43o'}
 F_2^{S(I),NLO}(x,t)=F_2^{S}(x,t_0)\frac{t}{t_0}\left( \frac{Y_{1}^{NLO}(x,t_0)}{ Y_{1}^{NLO}(x,t)}\right) \left[ \frac{X_{1}^{NLO}(x,t)-\frac{\beta}{t}}{X_{1}^{NLO}(x,t_0)-\frac{\beta}{t_0}} \right]
 \end{equation}
 %Put $X_2$ and $Y_2$ also\\

Similarly at NNLO, among the three functions $b_5(K,t,\sigma)$, $a_5(K,t,\sigma)$ and $c_5(K,t,\sigma)$  arising in eqn.(\ref{E5.37} )as given in Appendix B in the denominators of $X_1^{NNLO}$ and $Y_1^{NNLO}$ of  eqns.(\ref{D2}) and (\ref{D2'}) respectively for a given $K,t, \sigma$  the function $b_5(K,t,\sigma)$ is  smaller compared to $a_5(K,t,\sigma)$ and $c_5(K,t,\sigma)$. Under this assuumption $X_1^{NNLO}(x,t)$ and $Y_1^{NNLO}(x,t)$ are obtained as the following.
\begin{equation}
 \label{E5.43p}
 X_{1}^{NNLO}(x,t)=t^{\frac{b}{t}}e^{\frac{b}{t}-\frac{b^2(\ln t)^2+b^2+c}{2t^2}}\exp\left[ \frac{1}{a}\left( \frac{x}{a_5}-\frac{c_5}{a_5^{2}}\frac{x^2}{4}-\frac{c_5}{a_5^{2}}\frac{x^2}{2}\log{\frac{1}{x}}\right) \right] 
 \end{equation}
 \begin{multline}
 \label{E5.43q}
 Y_{1}^{NNLO}(x,t)=\exp[ x(a_6+d_6)-x^2\left( \frac{1}{4}\frac{a_6c_5}{a_5}-\frac{b_6}{2}+\frac{1}{4}\frac{c_5d_6}{a_5}\right)+\frac{x^3}{9}\frac{b_6c_5}{a_5} + x \log \frac{1}{x}d_6\\+x^2 \log\frac{1}{x}\left( -\frac{a_6c_5}{2a_5}-\frac{c_5d_6}{2a_5}\right) -x^3 \log\left( \frac{1}{x}\right)  \frac{b_6c_5}{3a_5}-x^2 \left( \log\frac{1}{x}\right) ^2 \frac{c_5d_6}{3a_5}] 
 \end{multline} 
 Where the coeffecients $a_5,c_5,a_6,b_6,d_6$ are as given in the Appendix B.\\
  Hence
 \begin{equation}
 \label{E5.43q'}
 F_2^{S(I),NNLO}(x,t)=F_2^{S}(x,t_0)\frac{t}{t_0}\left( \frac{Y_{1}^{NNLO}(x,t_0)}{ Y_{1}^{NNLO}(x,t)}\right) \left[ \frac{X_{1}^{NNLO}(x,t)-\frac{\beta}{t}}{X_{1}^{NNLO}(x,t_0)-\frac{\beta}{t_0}} \right]
 \end{equation}
 Eqns.(\ref{E5.26}) and (\ref{E5.37}) with the definitions of $X_1^{NLO}(x,t),Y_1^{NLO}(x,t) $ and $X_1^{NNLO}(x,t),Y_1^{NNLO}(x,t) $ as given in eqns.(\ref{E5.43o'}) and (\ref{E5.43q'}) are the equations to be tested in the following numerical section.

\section{Numerical Analysis}
\label{E}

\begin{figure*}
\resizebox{1.00\textwidth}{!}{%
  \includegraphics{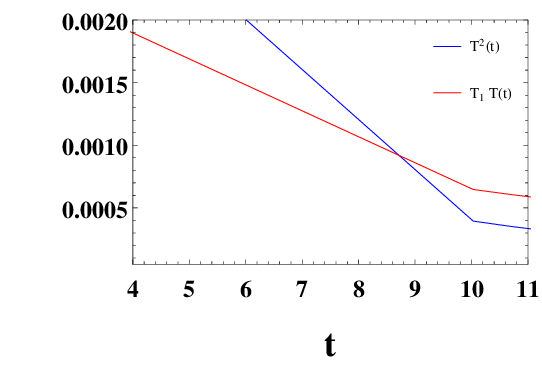}\quad \vspace{1cm} \includegraphics{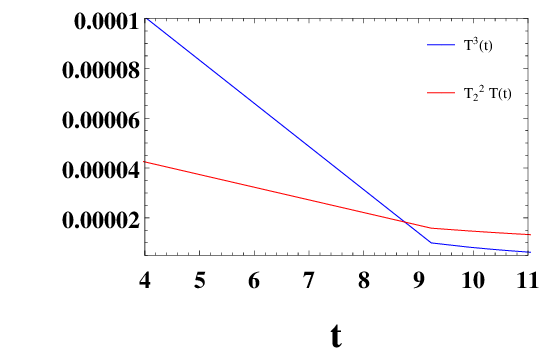}
}
\caption{ Plot of $T^2(t)^{NLO}$ and $T_1T(t)^{LO}$ (left fig.) and plot of $T^3(t)^{NNLO}$ and $T_2^2T(t)^{LO}$ (right fig.) versus $t$ for their correosponding $c^{NLO}$ and $c^{NNLO}$ taken to be unity.}
\label{fig:1}       % Give a unique label
\end{figure*}

Before testing the models with data and exact result let us first make some observation on the parameters $T_1$ and $T_2$ occurred in the present approach at NLO eqns.(\ref{E5.26})(\ref{E5.27}) and NNLO eqns.(\ref{E5.37})(\ref{E5.38}).\\
\textbf{NLO:}\\
%Let us now discuss the physical significance of the linearized parameter $T_1$ as occured in the formalism in eqns (\ref{E5.22}) and (\ref{E5.23}). 
From the above equations it is clear that $F_2^{S(I)NLO}$ %and $F_2^{S(II)NLO}$ 
contains a parameter $T_1$ $(<1)$ which does not occur at LO.  %In a sense this is one of the inherent limitations of the present approach beyond LO to be compared with the exact NLO solution of the DGLAP equations where such additional parameter does not occur. %However exact numerical value of $T_1$ has to be determined from data. 
It is therefore reasonable to relate $T_1$ as the average value of the QCD running coupling constant for the $Q^2$ range under study.

 \begin{equation}
 \label{E5.47}
 T_1 \propto \frac{<\alpha_s(t)^{NLO}>}{2\pi}=c^{NLO}\frac{1}{2\pi (t_2-t_1)}\int_{t_1}^{t_2} \alpha_s(t)^{NLO} dt
 \end{equation}
 The proportionality constant $c^{NLO}$ is first set as unity. Choosing $t_1$ and $t_2$ in the $Q$ range $2\leq Q \leq 100$ %we find $T_1=c^{NLO} \times 0.02704$. In eqn.(\ref{E5.47}) we take $c^{NLO}=1$ which gives  
 we obtain $T_1=0.02704$ from eqn.(\ref{E5.47}). To study the qualitative feature of the linearised relation, % we set $c^{NLO}=1$ 
  we plot $T^2(t)$ and $T_1T(t)$ as a function of $t$ as in fig.(\ref{fig:1})(left). The figure indicates that  the linearised approximation $T_1T(t)$  appears to be valid at around the intersection point of  $t=8.8$, corresponding to $Q^2=590$ GeV$^2$. %The approximation is  underestimated by the exact $T^2(t)$ whereas above it the linearised approximation over estimates the exact relation.\\% and only at higher end of $t$ values, it come closer.
 
  %obtain $\hat{T}_0$ to be 0.02704.\\\
\textbf{NNLO:}\\
% To fix the value of $\hat{T}_0$ , it is reasonable to identify $2\pi \hat{T}_0$ as the average value of the coupling constant over the $Q^2$ range under study.. For a range between $t_1=\ln \frac{Q_1^2}{\Lambda^2}$ to $t_2=\ln \frac{Q_2^2}{\Lambda^2}$,
% \begin{equation}
% \label{E4.26}
% \hat{T}_0=\frac{<\alpha_s(t)>}{2\pi}=\frac{1}{2\pi (t_2-t_1)}\int_{t_1}^{t_2} \alpha_s(t) dt
% \end{equation}
% Choosing $t_1$ and $t_2$ in the $Q$ range $2\leq Q \leq 100$ we obtain $\hat{T}_0$ to be 0.02704.
 Similarly for NNLO, the linearized parameter $T_2(<1)$  occurs in the formalism in eqns (\ref{E5.33}) and (\ref{E5.34}). %From the above equations it is clear that $F_2^{S(I),NNLO}$ and $F_2^{S(II),NNLO}$ contain two parameters $T_2$ $(<1)$. 
 For a range between $t_1=\ln \frac{Q_1^2}{\Lambda^2}$ to $t_2=\ln \frac{Q_2^2}{\Lambda^2}$,
 \begin{equation}
 \label{E5.48}
 T_2 \propto \frac{<\alpha_s(t)^{NNLO}>}{2\pi}=c^{NNLO} \frac{1}{ 2\pi(t_2-t_1)}\int_{t_1}^{t_2} \alpha_s(t)^{NNLO} dt
 \end{equation}
 Again the proportionality constant $c^{NNLO}$ is first set as unity. Choosing $t_1$ and $t_2$ in the $Q$ range $2\leq Q \leq 100$, %$T_2=c^{NNLO} \times 0.0247$ and 
% $T_2^2=(c^{NNLO}\times 0.0247)^2$. In eqn.(\ref{E5.48}) we take $c^{NNLO}=1$ which gives 
we obtain $T_2=0.0247$ from eqn.(\ref{E5.48}). To study the qualitative feature of the linearised relation, we plot fig.(\ref{fig:1})(right), and it shows that the linearisation approximation is valid at around small range of $t=8.8$ i.e $Q^2=590$ GeV$^2$.\\
We therefore first check the validity of our model around $Q^2=590$ GeV$^2$ %(i.e $t=8.5$) 
where the NLO and NNLO conditions are valid. 

In the previous work \cite{Lux2}, the range of validity of the model was $2\times10^{-6}\leq x\leq 1\times 10^{-3}$ and $2\leq Q\leq 100$ GeV for $\beta=-1.016, K=1.856,\sigma=0.227$. Using the same set of values for those parameters  we make our analysis at NLO and NNLO effects first with the obtained $T_1=0.02704$ and $T_2=0.0247$ respectively in the same $x$ and $Q^2$ range. \\

%So here we study only for $F_2^{S(I)}(x,t)$ and see how the result changes when including NLO and NNLO terms, incorporating the new parameters $T_1$ and $T_2$.
%For NLO putting the values of $\beta=-1.016, K=1.856,\sigma=0.227$ as in  \cite{Lux2}, $T_1=0.02704$ (NLO) and $N_f=4$, $a=\frac{2}{\beta_0}=0.24$, $b=\frac{\beta_1}{\beta_0^2}=0.739$ and $c=-1.573$ in eqns.(\ref{E5.40}) we find that our $F_2^{S(I),NLO}(x,t)$
  %grows very rapidly with $t$ resulting in a diagreement with the data \cite{HERA1.5,nnpdf}. 
  %Even in the case of NNLO using (\ref{E5.42'}) keeping the same values for the set of the above  parameters and $T_2=0.0247$ (NNLO) we obtain  $F_2^{S(I),NNLO}(x,t)$.	
    \begin{figure*}
\resizebox{0.6\textwidth}{!}{%
  \includegraphics{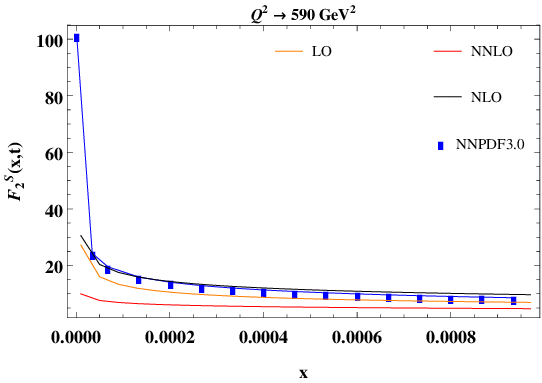}\quad \vspace{1cm} 
}
\caption{Plots of $F_2^{S(I),LO}$ and $F_2^{S(I),NLO}$, $F_2^{S(I),NNLO}$ as a function of $x$  for fixed $Q^2=590$ GeV$^2$ and comparison with the NNPDF3.0 data.}
\label{fig:3}       % Give a unique label
\end{figure*}
\begin{figure*}
\resizebox{0.6\textwidth}{!}{%
  \includegraphics{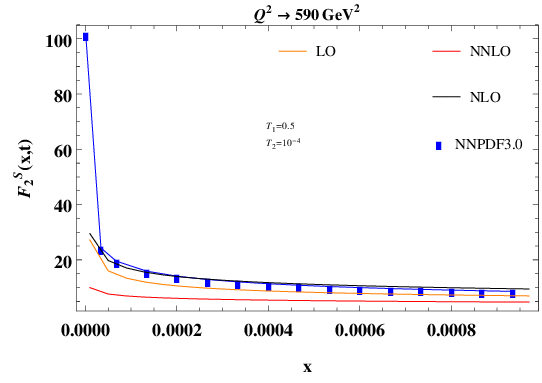}\quad \vspace{1cm} 
}
\caption{Plots of $F_2^{S(I),LO}$ and $F_2^{S(I),NLO}$, $F_2^{S(I),NNLO}$ as a function of $x$  for fixed $Q^2=590$ GeV$^2$ using $T_1=0.5$ and $T_2= 10^{-4}$ and comparison with the NNPDF3.0 data.}
\label{fig:3i}       % Give a unique label
\end{figure*}

%Fig.(\ref{fig:3}) We plot  comparison with the data needs to be done.)
In Fig.{\ref{fig:3}} we plot the singlet structure Functions $F_2^{S(I),LO}$ and $F_2^{S(I),NLO}$, $F_2^{S(I),NNLO}$   for a range of small $2\times10^{-6}\leq x\leq 1\times 10^{-3}$ and at fixed $Q^2=590$ GeV$^2$ and then compare with the NNPDF3.0 data \cite{nnpdf}. We observe that the LO as well as the NLO effects are much closer to data. But when NNLO effects are added the prediction comes below the data.
For instance it is observed that varying the values of $T_1$ and  $T_2$ bring the prediction closer to data. As an illustration we show  for  $T_1=0.5$ and $T_2= 10^{-4}$ with the corresponding values of $C^{NLO}= 18.491$ and $C^{NNLO}=0.004 $ as in fig.(\ref{fig:3i}). At those values of $T_1$ and $T_2$ the NLO prediction shows a high consistency with the data but we do not observe any significant rise in the evolution for NNLO correction. which interprets that the parameter $T_2$ is having a very minimal contribution in the evolution for NNLO correction.\\%compared to when considering the NNLO effects. \\
\begin{figure*}
\resizebox{1.00\textwidth}{!}{%
  \includegraphics{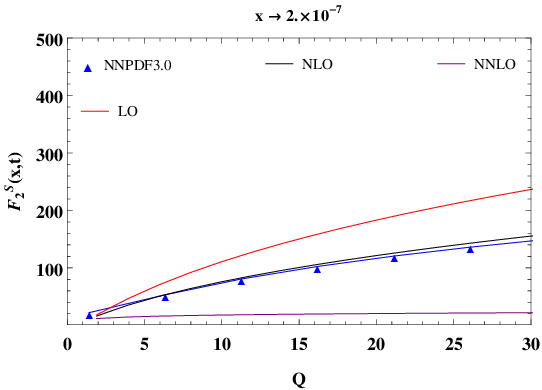}\quad \vspace{1cm} \includegraphics{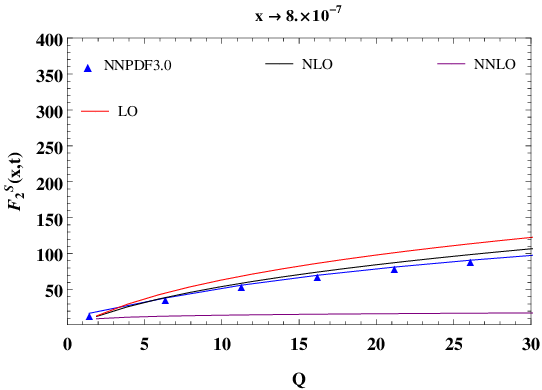}
  \quad \vspace{1cm} \includegraphics{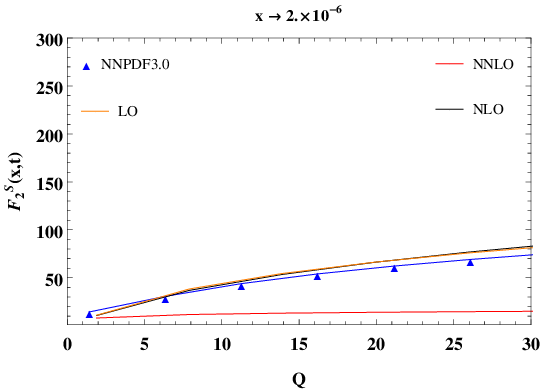}
}
 % \resizebox{1.00\textwidth}{!}{%
  %\includegraphics{4fig4.eps}\quad \vspace{1cm} \includegraphics{4fig5.eps}
  %\quad \vspace{1cm} \includegraphics{Ba9.eps}
%}
\resizebox{1.00\textwidth}{!}{%
  \includegraphics{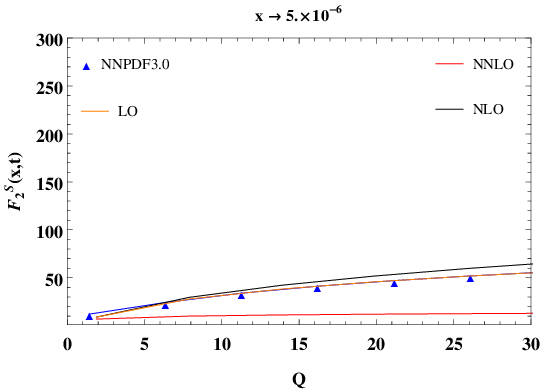}\quad \vspace{1cm} \includegraphics{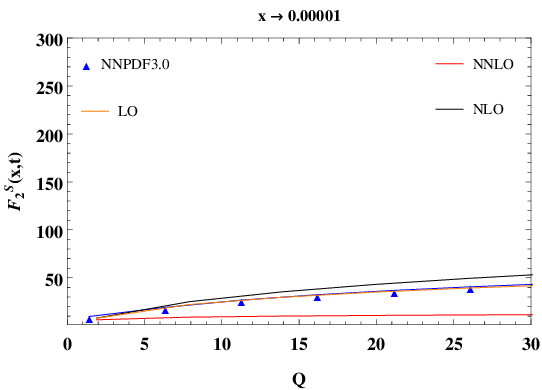} \includegraphics{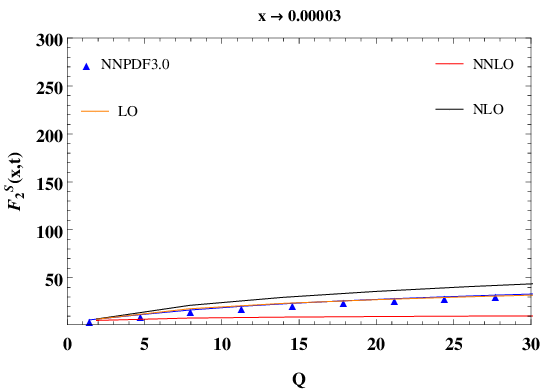}
}

\caption{Plots of $F_2^{S(I),LO}$ and $F_2^{S(I),NLO}$, $F_2^{S(I),NNLO}$ as a function of $Q$  for fixed $x$ and comparison with the NNPDF3.0 \cite{nnpdf}.}
\label{fig:4}       % Give a unique label
\end{figure*}
%\begin{figure*}
%%\resizebox{1.00\textwidth}{!}{%
%  %\includegraphics{4fig1.eps}\quad \vspace{1cm} \includegraphics{4fig3.eps}
%  %\quad \vspace{1cm} \includegraphics{Ba6.eps}
% % }
% % \resizebox{1.00\textwidth}{!}{%
%  %\includegraphics{4fig4.eps}\quad \vspace{1cm} \includegraphics{4fig5.eps}
%  %\quad \vspace{1cm} \includegraphics{Ba9.eps}
%%}
%\resizebox{1.00\textwidth}{!}{%
%  \includegraphics{5fig1.eps}\quad \vspace{1cm} \includegraphics{5fig2.eps}
%}
%\resizebox{1.00\textwidth}{!}{%
%  \includegraphics{5fig3.eps}\quad \vspace{1cm} \includegraphics{5fig4.eps}
%}
%
%
%\caption{Plots of $F_2^{S(I),LO}$ and $F_2^{S(I),NLO}$, $F_2^{S(I),NNLO}$ as a function of $Q$  for fixed $x$ and comparison with the NNPDF3.0 \cite{nnpdf}.}
%\label{fig:4}       % Give a unique label
%\end{figure*}
\\
Fig.{\ref{fig:4}} represents the results of $Q$ evolution of the singlet structure Functions $F_2^{S(I),LO}$ and $F_2^{S(I),NLO}$, $F_2^{S(I),NNLO}$ for a few values of small $x$. Using our initial set of parameters $T_1=0.02704$ and $T_2=0.0247$ %obtained from averaging over the integrated coupling constant 
in our models $F_2^{S(I),NLO}$ (eqn.\ref{E5.26}) and $F_2^{S(I),NNLO}$ (eqn.\ref{E5.37}), we compare with the NNPDF3.0 data \cite{nnpdf}.  We observe that our model upto NLO is valid but when NNLO corrections are taken then the structure function deviates much below the data as well as with the LO and NLO result. This shows that our small $x$ approximations in NLO model is validated than those in our NNLO model. %Phenomenological observation  allow only model $F_2^{S(I),NLO}$  and not $F_2^{S(I),NNLO}$  to be compatible with the available data in the small $x$ range between $1\times 10^{-8}\leq x \leq 8\times 10^{-6}$ and $Q^2\geq 590$ GeV$^2$ (For $Q^2=$ and $x=$ 
%the ratio of $\frac{F_2^{NLO}}{F_2{LO}}\simeq... $ and $\frac{F_2^{NNLO}}{F_2{LO}}\simeq ....$). This shows that our small $x$ approximations in NLO model is validated than those in our NNLO model. %Our model of NNLO is valid for much larger $x$ range.

\section{Conclusion}
\label{F}
In this paper, we obtained the approximate analytical expressions for the singlet structure functions both at NLO and NNLO from the Taylor approximated coupled DGLAP equations assuming the %$t$ dependence 
adhoc $t$ dependent relation between the singlet and the gluon distribution  consistent with the analysis of Lopez and Yndurain \cite{ly,ne}. At LO the phenomenological analysis is done with the three sets of parameters $\beta=-1.016, K=1.856,\sigma=0.227$  while for NLO and NNLO additional parameters $T_1=0.02704$ and $T_2=0.0247 $ respectively are needed. Our analysis are done around $Q^2=590$ GeV$^2$ which corresponds to our approximation where our higher order approximation is valid. We make the analysis of both the variation of structure function with $Q$ and $x$.   \\
At the extreme limit of $T_2=0$ theoretically we recover the NLO result (as can be observed from eqn.(\ref{E5.33}) and (\ref{E5.34})) but we do not obtain it from the analytical expression (eqn.(\ref{E5.43p}) and (\ref{E5.43q})), which may predict that the various approximations assumed in solving the DGLAP equations upto NNLO may not be all justified.

Let us now conclude this work with a few comments regarding its limitation. There are several approximations which are used to analytically solve the DGLAP equations at small $x$ upto NNLO with valid degree of justification.\\
(i) The PDFs are expanded at small $x$ (eqn.\ref{E5.7'}) which has inherent theoretical limitations. In DGLAP evolutions, the integrals extends from  0 to 1. Thus at small $x$ one cannot neglect in principle the medium/large $x$ behaviour of the Parton Distribution functions. \\
(ii) Equation(\ref{E5.12}) relating the singlet and the Gluon distribution assumes the $x$ dependence of both the distribution to be the same upto a proportionality factor, which may not be true in general. At small scale $(Q^2)$ the standard PDF available in current literarture Ref.\cite{JHEP01} suggest such difference. Even in ultra small $x$ and ultra high $Q^2$ double asymptotic scaling limit suggest different $x$ dependence for the singlet and gluon distribution. Specifically at double asymptotic limit (ultra small $x$ and ultra large $Q^2$) the corresponding gluon and the singlet distribution have different $x$ dependence in general. \cite{RDBall, DKCINT,Lux2}. Thus the present assumption appears to be approximately true at limited $x$ and $Q^2$ as has been observed in the present work.\\
(iii) In analytical solutions beyond LO, we need to linearise the $t-$ dependence of the coupling constant $\alpha^{NLO}$ (eqn.(\ref{E5.5})) and $\alpha^{NNLO}$ (eqn.(\ref{E5.6})). It introduces the parameter $T_1$ and $T_2$ at NLO and NNLO respectively, which is the inherent limitation of the present analytical approach as had been noted  in earlier communication \cite{PKD,Mayuri}. %The best values of $T_1$ and $T_2$ are found to be ...respectively.
 Such approximations are however found to be true in the limited $Q^2$ around 590 GeV$^2$.\\

\section*{Appendix A:}
The expressions for the functions $A_i^S(x,t), B_i^S(x,t), C_i^S(x,t),D_i^S(x,t), L_i^S(x,t), M_i^S(x,t)$ where $i=1,2$ involving the integral representation containing the splitting functions as occured in the eqns.(\ref{E5.13}), (\ref{E5.14}) and (\ref{E5.30}), (\ref{E5.31})   are as follows. $A_i^S(x,t), B_i^S(x,t), C_i^S(x,t),D_i^S(x,t)$ are same in both the cases of NLO and NNLO.
\begin{eqnarray}
\label{B1.1}
A_1^S(x,t)=\int_x^1 dz\left[ P_{qq}^{LO}(x)+2 N_fP_{qg}^{LO}(x) Kt^{\sigma}\right] \\\label{B1.2}B_1^S(x,t)=\int_x^1 dz\left[P_{qq}^{LO}(x)\frac{x(1-z)}{z}+2 N_fP_{qg}^{LO}(x)\frac{x(1-z)}{z} Kt^{\sigma}\right] \\\label{B1.3}C_1^S(x,t)=\int_x^1 dz\left[ P_{qq}^{NLO}(x)+2 N_fP_{qg}^{NLO}(x) Kt^{\sigma}\right] \\\label{B1.4}D_1^S(x,t)=\int_x^1 dz\left[P_{qq}^{NLO}(x)\frac{x(1-z)}{z}+2 N_fP_{qg}^{NLO}(x)\frac{x(1-z)}{z} Kt^{\sigma}\right]\\\label{B2.1}
A_2^S(x,t)=\frac{1}{K t^{\sigma}}\int_x^1 dz\left[ P_{gg}^{LO}(x) Kt^{\sigma}+P_{gq}^{LO}(x)\right] \\\label{B2.2}B_2^S(x,t)=\frac{1}{K t^{\sigma}}\int_x^1 dz\left[P_{gg}^{LO}(x)\frac{x(1-z)}{z}Kt^{\sigma}+P_{gq}^{LO}(x)\frac{x(1-z)}{z}\right] \\\label{B2.3}C_2^S(x,t)=\frac{1}{K t^{\sigma}}\int_x^1 dz\left[ P_{gg}^{NLO}(x)Kt^{\sigma}+P_{gq}^{NLO}(x)\right] \\\label{B2.4}D_2^S(x,t)=\frac{1}{K t^{\sigma}}\int_x^1 dz\left[P_{gg}^{NLO}(x)\frac{x(1-z)}{z}Kt^{\sigma}+P_{gq}^{NLO}(x)\frac{x(1-z)}{z}\right]
\end{eqnarray}
%\begin{equation}
%A_2^S(x,t)=\frac{1}{K t^{\sigma}}\int_x^1 dz\left[ P_{gg}^{LO}(x) Kt^{\sigma}+P_{gq}^{LO}(x)\right]
%\end{equation}
%\begin{equation}
%B_2^S(x,t)=\frac{1}{K t^{\sigma}}\int_x^1 dz\left[P_{gg}^{LO}(x)\frac{x(1-z)}{z}Kt^{\sigma}+P_{gq}^{LO}(x)\frac{x(1-z)}{z}\right] 
%\end{equation}
And
\begin{eqnarray}
\label{B1.5}L_1^S(x,t)=\int_x^1 dz\left[ P_{qq}^{NNLO}(x)+2 N_fP_{qg}^{NNLO}(x) Kt^{\sigma}\right] \\\label{B1.6}M_1^S(x,t)=\int_x^1 dz\left[P_{qq}^{NNLO}(x)\frac{x(1-z)}{z}+2 N_fP_{qg}^{NNLO}(x)\frac{x(1-z)}{z} Kt^{\sigma}\right]\\
\label{B2.5}L_2^S(x,t)=\frac{1}{K t^{\sigma}}\int_x^1 dz\left[ P_{gg}^{NNLO}(x)Kt^{\sigma}+P_{gq}^{NNLO}(x)\right] \\\label{B2.6}M_2^S(x,t)=\frac{1}{K t^{\sigma}}\int_x^1 dz\left[P_{gg}^{NNLO}(x)\frac{x(1-z)}{z}Kt^{\sigma}+P_{gq}^{NNLO}(x)\frac{x(1-z)}{z}\right]
\end{eqnarray}

\section*{Appendix B:}
The coefficients $a_1(K,t,\sigma), b_1(K,t,\sigma),c_1(K,t,\sigma), a_2(K,t,\sigma),b_2(K,t,\sigma), d_2(K,t,\sigma),a_3(K,t,\sigma),b_3(K,t,\sigma),\\c_3(K,t,\sigma),d_3(K,t,\sigma),a_4(K,t,\sigma),b_4(K,t,\sigma),d_4(K,t,\sigma)$ as given in eqns.(\ref{D1})-(\ref{D1'''}) and the coeffecients $a_5(K,t,\sigma), b_5(K,t,\sigma),c_5(K,t,\sigma), a_6(K,t,\sigma),b_6(K,t,\sigma), d_6(K,t,\sigma),a_7(K,t,\sigma),b_7(K,t,\sigma),c_7(K,t,\sigma),\\d_7(K,t,\sigma),a_8(K,t,\sigma),b_8(K,t,\sigma),d_8(K,t,\sigma)$ as given in eqns.(\ref{D2})-(\ref{D2'''}) are listed below.
\begin{eqnarray}
\label{D6.1'}
a_1^{}(K,t,\sigma, T_1)=T_1(\frac{80}{27}N_f+\frac{20}{3}Kt^{\sigma})\nonumber\\ \\\label{D6.1'}
a_2^{}(K,t,\sigma, T_1)=\frac{8}{3}+\frac{2 N_fKt^{\sigma}}{3}+T_1.(-435.136-\frac{384217 N_f}{810}+K t^{\sigma}(\frac{16 \pi^2+649}{216}N_f+\frac{1219597}{1080}))\nonumber\\+T_1(-3414+\frac{23}{6} K t^{\sigma})\ln(2)\nonumber\\ \\\label{D6.2'}a_3^{}(K,t,\sigma, T_1)=(6+\frac{8}{3Kt^{\sigma}})+T_1(\frac{3843}{48}-\frac{61}{9}N_f-3\pi^2+\frac{1}{Kt^{\sigma}}(\frac{517}{18}-\frac{4}{3}\pi^2-\frac{80}{27}N_f)) \nonumber\\ \\\label{D6.2}
a_4^{}(K,t,\sigma, T_1)=(\frac{11}{12}-\frac{N_f}{3}-11)-\frac{12}{Kt^{\sigma}}-T_1(\frac{20035}{3601}+\frac{7 \pi^2}{2}-\frac{679705.51}{1320}+\frac{1}{Kt^{\sigma}}(\frac{80}{27}N_f+\frac{11 \pi^2}{3}\nonumber\\+\frac{2317}{864}+\frac{103837.66}{81})+T_1(\frac{45}{2})\\\label{D6.2a}
a_5^{}(K,t,\sigma, T_2)=T_2(\frac{80 N_f}{27}+\frac{20Kt^{\sigma}}{3})+T_2^2(-19666.936+38600.48 Kt^{\sigma})  \\\label{D6.3}
a_6^{}(K,t,\sigma, T_2)=\frac{8}{3}+\frac{2 N_fKt^{\sigma}}{3}+T_2(-435.136-\frac{384217 N_f}{810})+T_2(-3414+\frac{23}{6}Kt^{\sigma})\ln(2)\nonumber\\+T_2^2(4061.071-3921.587\times 8Kt^{\sigma})\\ \label{D6.3a}
a_7^{}(K,t,\sigma, T_2)=(6+\frac{8}{3Kt^{\sigma}})+T_2\left( \frac{3843}{48}-\frac{61 N_f}{9}-3 \pi^2 +\frac{1}{Kt^{\sigma}}(\frac{517}{18}-\frac{4 \pi^2}{3}-\frac{80 N_f}{27}) \right)\nonumber\\ +T_2^2(135983.174+\frac{3240.749}{Kt^{\sigma}})\\\label{D6.4}
 a_8^{}(K,t,\sigma, T_1)=(\frac{11}{12}-\frac{N_f}{3}-11)-\frac{12}{Kt^{\sigma}}-T_1(\frac{20035}{3601}+\frac{7 \pi^2}{2}-\frac{679705.51}{1320}+\frac{1}{Kt^{\sigma}}(\frac{80}{27}N_f+\frac{11 \pi^2}{3}\nonumber\\+\frac{2317}{864}+\frac{103837.66}{81})+T_1(\frac{45}{2})+T_2^2(-87979.883-\frac{28540.98}{Kt^{\sigma}})\nonumber\\ \\\label{D6.5'} b_1^{}(K,t,\sigma, T_1)=-\frac{5}{3}N_fKt^{\sigma}+T_1(\frac{416151}{324}N_f-\frac{627740.73}{5832}+Kt^{\sigma}(\frac{1659953}{2160}+\frac{\pi^2}{27})+(-\frac{204}{81}-\frac{883}{4}Kt^{\sigma})\ln(2))\nonumber\\ \\\label{D6.5}
 b_2^{}(K,t,\sigma, T_1)=(\frac{4}{3}- N_f K t^{\sigma})+T_1(\frac{-16791.29}{12}+\frac{3593 N_f}{108}+\frac{125 K t^{\sigma}}{16})\nonumber\\ \\\label{D6.6}
 b_4^{}(K,t,\sigma, T_1)=\frac{1}{Kt^{\sigma}}(12Kt^{\sigma}+16)+T_1(\frac{679}{108}N_f-\frac{2305}{32}-4\pi^2-\frac{1}{Kt^{\sigma}}(\frac{79649}{81}-\frac{4\pi^2}{3}-\frac{128}{27}N_f))\nonumber\\\label{D6.6a}
 b_5^{}(K,t,\sigma, T_2)= -\frac{5}{3}N_fKt^{\sigma}+T_2\left( \frac{416151N_f}{324}-\frac{627740.73}{5832}+Kt^{\sigma}(\frac{1659953}{2160}+\frac{\pi^2}{27})\right) \nonumber\\+T_2\left( -\frac{204}{81}\ln(2)-\frac{883}{4}Kt^{\sigma}\ln(2)\right) +T_2^2\left( -12075.005-329142.4Kt^{\sigma}\right) \\\label{D6.7}
 b_6^{}(K,t,\sigma, T_2)=(\frac{4}{3}- N_f K t^{\sigma})+T_2(\frac{-16791.29}{12}+\frac{3593 N_f}{108}+\frac{125 K t^{\sigma}}{16})+T_2^2(-2143.669-22467.4\nonumber\\\times 8Kt^{\sigma})\\ \label{D6.7a}
 b_7^{}(K,t,\sigma, T_2)=(8+\frac{2}{3Kt^{\sigma}})+T_2\left( \frac{989}{108}N_f+4\pi^2+\frac{2557}{32}-\frac{20}{Kt^{\sigma}}\right) +T_2\left( \frac{1}{Kt^{\sigma}}(\frac{5104}{81}+4\pi^2+\frac{208}{27}N_f)\right) \nonumber\\+T_2^2(-139453.698-\frac{22436.898}{Kt^{\sigma}})
 \end{eqnarray}
 \begin{eqnarray}
 \label{D6.7'}
 b_8^{}(K,t,\sigma, T_2)=\frac{1}{Kt^{\sigma}}(12Kt^{\sigma}+16)+T_2(\frac{679}{108}N_f-\frac{2305}{32}-4\pi^2-\frac{1}{Kt^{\sigma}}(\frac{79649}{81}-\frac{4\pi^2}{3}-\frac{128}{27}N_f))\nonumber\\+T_2^2(49919.486+\frac{19159.5}{Kt^{\sigma}})\nonumber\\ \\\label{D6.8}c_1^{}(K,t,\sigma, T_1)=T_1 (\frac{39634.98}{144}-\frac{158355}{4374}N_f+Kt^{\sigma}(\frac{395}{216}N_f-\frac{1533}{144}))\nonumber\\ \\\label{D6.9}c_5^{}(K,t,\sigma, T_2)=T_2(\frac{39634.98}{144}-\frac{158355}{4374}N_f+Kt^{\sigma}(\frac{395}{216}N_f-\frac{1533}{144}))+T_2^2(3866.215\nonumber\\+218339.68Kt^{\sigma})\nonumber\\ \\\label{D6.9'}c_7^{}(K,t,\sigma, T_2)=(-12+\frac{16}{3Kt^{\sigma}})-T_2\left\lbrace \frac{989}{108}N_f+4\pi^2+\frac{2557}{32}-\frac{20}{Kt^{\sigma}}\right\rbrace + \nonumber\\T_2\left\lbrace \frac{1}{Kt^{\sigma}}\left( \frac{5104}{81}+4\pi^2+\frac{208}{27}N_f\right)   \right\rbrace+T_2^2\left\lbrace -139453.698-\frac{22436.898}{Kt^{\sigma}}\right\rbrace \\\label{D6.10}
 d_2^{}(K,t,\sigma, T_1)=T_1(\frac{80}{27}N_f+\frac{20}{3}Kt^{\sigma})\nonumber\\\label{D6.11}
 d_4^{}(K,t,\sigma, T_1)=\frac{1}{KT^{\sigma}}(KT^{\sigma}+\frac{8}{3})+T_1\left( \frac{3843}{48}-\frac{61}{9}N_f-3\pi ^2+\frac{1}{KT^{\sigma}}(\frac{517}{18}-\frac{4\pi ^2}{3}-\frac{80}{27}N_f)\right)  \\\label{D6.12}
 d_6^{}(K,t,\sigma, T_2)=T_2(\frac{80}{27}N_f+\frac{20}{3}Kt^{\sigma})+T_2^2(-1908740+4825.06\times 8Kt^{\sigma})\nonumber\\ \\
\label{D6.13}
 d_8^{}(K,t,\sigma, T_2)=\frac{1}{KT^{\sigma}}(KT^{\sigma}+\frac{8}{3})+T_2\left( \frac{3843}{48}-\frac{61}{9}N_f-3\pi ^2+\frac{1}{KT^{\sigma}}(\frac{517}{18}-\frac{4\pi ^2}{3}-\frac{80}{27}N_f)\right) \nonumber\\+T_2^2(135983.313-\frac{3240.75}{Kt^{\sigma}})\nonumber\\ \\\label{D6.14}
 d_3^{}(K,t,\sigma, T_1)=T_1 \frac{45}{2}\nonumber\\ \\\label{D6.15}
 d_7^{}(K,t,\sigma, T_2)=T_2 \frac{45}{2}\nonumber\\
\end{eqnarray}


\begin{thebibliography}{10}
 \bibitem{Luxmi} Luxmi Machahari and D. K. Choudhury, Eur. Phys. J. A. (2018) \textbf{54}: 69
 \bibitem{Lux2} Luxmi Machahari and D. K. Choudhury, Commun. Theor. Phys.\textbf{71} (2019) 56-66.
 \bibitem{gl} V. N. Gribov and L. N. Lipatov, Sov. J. Nucl. Phys., \textbf{438}, (1972) 15.
\bibitem{l} L. N. Lipatov, Sov. J. Nucl. Phys., \textbf{20}, (1975) 94.
\bibitem{d} Yu. L. Dokshitzer, Sov. Phys. JETP, \textbf{46}, (1977) 641.
\bibitem{ap} G. Altarelli and G. Parisi, Nucl. Phys., B\textbf{126}, (1977) 298 .
\bibitem{h1}  H1 and ZEUS Collaborations (Aaron, F.D. et al.) JHEP \textbf{1001}, (2010) 109.
\bibitem{NLO2} R.K.Ellis , W.J.Stirling and B.R.Webber, QCD and Collider Physics(Cambridge University Press,1996).
\bibitem{NLO1} W. Furmanski and B. Petronzio, Phys. Lett B\textbf{97}, 437 (1980);\\ G. Curci, W. Furmanski and R. Petronzio, Nucl. Phys. B\textbf{175}, 27 (1980);\\ R.K.Ellis , W.J.Stirling and B.R.Webber, QCD and Collider Physics(Cambridge University Press,1996).
\bibitem{Moch1} S.Moch, J.Vermaseren and A.Vogt, Nucl.Phys.B \textbf{688}, 101(2004).
 \bibitem{Moch2} S.Moch, J.Vermaseren and A.Vogt, Nucl.Phys.B \textbf{691}, 129(2004).
 \bibitem{Retey} A.Retey, J.Vermaseren , Nucl.Phys.B \textbf{604}, 281(2001).
 \bibitem{ly} C. Lopez and F. J. Yndurain, Nucl. Phys. B \textbf{171}, (1980) 231.
 \bibitem{s} I. Sneddon, \textit{Elements of Partial Differential Equations} (M.Graw Hill, NewYork,1957) p131.
 \bibitem{LO1} P. D. Collins, An introduction to Regge theory an high-energy Physics (Cambridge University Press, Cambridge 1997) Cambridge; M. Bertini et al., Rivista del Nuovo Cmento 19, 1 (1996)
\bibitem{LO2} R. G. Roberts, The Structure of the Proton (Cambridge University Press 1990).
 \bibitem{G.R} G.R.Boroun and B.Rezaei, Eur. Phys. J. C \textbf{73} (2013)2412: arXiv: 1402.0164 [hep-ph].
 \bibitem{coupling} B. G. Shaikhatdenov, A. V. Kotikov, V. G. Krivokhizhin, and G. Parente, Phys.Rev.D\textbf{81},034008(2010).
 \bibitem{Borou} G.R. Boroun, Eur. Phys. J. A 50, 69 (2014).
 \bibitem{Akbari} A. Jahan and D.K. Choudhury, Commun.Theor. Phys. \textbf{61}(2014) 654-658.
\bibitem{ne} D.K. Choudhury and Neelakshi N.K. Borah, Physical Review D \textbf{95}, (2017) 014002.
\bibitem{Atri2} A Deshmukhya and D. K. Choudhury Proc. of the 2nd Regional Conf. on Phys. Research in North-East, Guwahati, India, 34 October (2000).
\bibitem{PKD3} D. K. Choudhury and P. K. Dhar, Indian J. Phys. \textbf{81} 819 (2007).
\bibitem{PKD4} D. K. Choudhury and P. K. Dhar, Indian J. Phys. \textbf{81} 259 (2007).
\bibitem{Saiful1} Saiful Islam and  D. K. Choudhury, Eur. Phys. J. C \textbf{72}, (2012) 2257.
\bibitem{JK2} M. Devee, R. Baishya and J. K. Sarma, Eur. Phys. J. C \textbf{72}, (2012) 2036 .
\bibitem{Ruju} A. De Rujula, S. L. Glashow, H. D. Politzer, S. B. Tieman, F. Wilczek and A.Zee, Phys. Rev. D \textbf{10} (1974) 1649.
\bibitem{Duhr} Claude Duhr, Nuclear Physics B Proceedings Supplement 00 (2014) 1-8.

\bibitem{HERA1.5} HERAPDF1.5: https : //www.desy.de/h1zeus/combined r esults/herapdf table/.
 \bibitem{nnpdf} R. D. Ball et al., Journal of High Energy Physics \textbf{04} (2015) 040.
\bibitem{Boroun5} G. R. Boroun and B. Rezaei, Eur. Phys. J. C \textbf{73}, (2013) 2412.
\bibitem{Block} Martin M. Block, Loyal Durand and Douglas W. McKay, Phys. Rev. D\textbf{77}, (2008) 094003.
\bibitem{Laplace1} S. Weinzierl, arXiv:hep-ph/0203112.
\bibitem{Soft1} A.Donnachie and P.V.Landshoff,Z. Phys. C\textbf{61}, 139 (1994); Phys. Lett. B\textbf{518}, 63 (2001)
\bibitem{Hard1} A.Donnachie and P.V.Landshoff, Phys. Lett. B\textbf{550}, 160(2002); R. D. Ball and  P.V.Landshoff, arXiv:hep-ph/9912445. %\\
P.Desgrolard, M. Giffon, E.Martynov and E .Predazzi, Eur. Phys. J. C\textbf{18}, 555(2001);\\
P.Desgrolard, M. Giffon and E.Martynov, Eur. Phys. J. C\textbf{7}, 655(1999);\\
A.D.Martin, M.G.Ryskin and G.Watt, arXiv:hep-ph/0406225.
\bibitem{JHEP01} H1 and ZEUS Collaboration, Journal of High Energy Physics \textbf{01} (2010) 109.
\bibitem{RDBall} R. D. Ball and S. Forte, Phys. Lett. B \textbf{336}, 77(1994), CERN-THH-7331/94.
\bibitem{DKCINT} D. K. Choudhury, International Centre for Theoretical Physics, Trieste, Italy IC/94/324
\bibitem{PKD} D.K. Choudhury and Pijush. Kanti. Dhar, Indian J. Phys. \textbf{83}, (2009) 1199. 
\bibitem{Mayuri} Mayuri Devee, R. Baishya and J. K. Sarma,  Eur. Phys. J. C (2012) \textbf{72}: 2036.
 
 
% \bibitem{van1} W.L. van Neerven, A.Vogt, Nucl.Phys.B \textbf{588}, 345(2000).
% \bibitem{van2} W.L. van Neerven, A.Vogt, Nucl.Phys.B \textbf{568}, 263(2000).
% 



%\bibitem{Block5} Martin M. Block, Loyal Durand, Phuoc Ha, Douglas W. Mckay, Eur. Phys. J. C \textbf{69}, (2010) 425.





\end{thebibliography}
\end{document}